\documentstyle[preprint,aps]{revtex}
\begin{document}
\preprint{KIAS-P99108}
%\preprint{KNU-H9910}
\def\a{\alpha}
\def\b{\beta}
\def\p{\partial}
\def\m{\mu}
\def\n{\nu}
\def\s{\sigma}
\def\t{\tau}
\def\half{\frac{1}{2}}
\def\hatt{{\hat t}}
\def\hatx{{\hat x}}
\def\hatth{{\hat \theta}}
\def\hatta{{\hat \tau}}
\def\hatrh{{\hat \rho}}
\def\hatva{{\hat \varphi}}
\def\p{\partial}
\def\nn{\nonumber}
\def\beq{\begin{eqnarray}}
\def\eeq{\end{eqnarray}}

\title{Canonical Quantization of Open String and \\
Noncommutative Geometry}
\author{Taejin Lee
%$^{1}$
\thanks{taejin@cc.kangwon.ac.kr}}
\address{{\it 
 %${}^{1}$
 Department of Physics, Kangwon National University, 
 Chuncheon 200-701, Korea}}
\date{\today}
\maketitle
\begin{abstract}
We perform canonical quantization of open strings in the $D$-brane
background with a $B$-field. Treating the mixed boundary condition
as a primary constraint, we get a set of secondary constraints. Then
these constraints are shown to be equivalent to orbifold 
conditions to be imposed on normal string modes. These orbifold conditions
are a generalization of the familiar orbifold conditions which arise when 
we describe open strings in terms of closed strings. Solving the 
constraints explicitly, we obtain a simple Hamiltonian for the open
string, which reveals the nature of noncommutativity transparently.

\end{abstract}


\narrowtext

\section{Introduction}
The open string gives rise to the noncommutative geometry\cite{nc} for
the $D$-brane with a NS-NS $B$-field. The $D$-brane dynamics
is described by Yang-Mills gauge fields on noncommutative 
space-time. This point was implied in the work of Connes, Douglas,
and Schwarz\cite{conn} on the Matrix $M$-model\cite{mth} 
compactified on a torus in an appropriate limit.
Subsequently, more direct approaches to the noncommutative geometry in 
the string theories were taken in refs.\cite{open,biga,ard1,ho1},
where the open string dynamics in the $D$-brane background are
studied. The various aspects of the noncommutative Yang-Mills gauge
theories and their implications in the string theories were discussed
extensively in a recent work of Seiberg and Witten\cite{seib}.
In particular the equivalence of the ordinary gauge fields and 
the noncommutative gauge fields has been proposed and checked by 
comparing the ordinary Dirac-Born-Infeld theory with its noncommutative 
counterpart for the $D$-brane. 

In order to explore further the noncommutative geometry in 
the string and its nonperturbative effects,
we may need to develop the string field theory based on the
noncommutative algebra. Bigatti and Susskind\cite{biga} also 
discussed recently 
relevance of the noncommutative geometry in the light cone quantization
of open strings attached to $D$-brane, which can be easily extended to
the light cone string field theory. In this respect it is important 
to perform canonical quantization of the open string theory 
in the background of $D$-brane, which will be a stepping stone toward
the second quantized theory. Quantization of the open strings 
in the presence of $D$-branes with a $B$-field has been already
discussed in the literature. In ref.\cite{ard2} it was pointed out that the 
nontrivial boundary condition in the presence of the $B$-field 
modify the canonical commutation relations and
leads to the noncommutativity on $D$-brane worldvolume.
This point was elaborated further subsequently by Chu and Ho\cite{ho2}, 
as they examine the simplectic form obtained in terms of the mode expansion
of the classical solutions. However, there are some discrepancy
between two works. To resolve the discrepancy the authors of both works
perform the canonical quantization, treating the mixed boundary condition
as a primary constraint and employing the Dirac's quantization method.
Nevertheless, the discrepancy still remains. The purpose of this paper
is to carry out the canonical quantization of the open string in the
$D$-brane background with some rigor and to clarify the related 
problems. In the course we will be able to confirm some of the results
obtained in \cite{seib} by using the conformal field theory.

\section{Open String in the Background of $D$-Brane}

The bosonic part of the classical action for an open 
string ending on a $Dp$-brane with 
a $B$-field is given by
\begin{eqnarray} \label{action}
I = \frac{1}{4\pi\alpha^\prime}\int_M d^2 \xi 
\left[G_{\m\n} \sqrt{-h} h^{\alpha\beta} \frac{\p X^\m}{\p \xi^\a}
\frac{\p X^\n}{\p \xi^\b} + B_{ij} \epsilon^{\a\b}
\frac{\p X^i}{\p \xi^\a}\frac{\p X^j}{\p \xi^\b} \right]
\end{eqnarray}
where $\m = 0, 1, \dots, 9$ and $i=0, 1, \dots, p$.
Here we consider a simple flat background first: 
$G_{\m\n}= \eta_{\m\n}$, $H=dB=0$. Extension to a more general 
background will be discussed later.
If the $U(1)$ gauge field on the $Dp$-brane is present,
$(B_{ij}+F_{ij})$ replaces $B_{ij}$ in the action Eq.(\ref{action}).
For simplicity we set $2\pi\alpha^\prime =1$ and restore it when
necessary.

Choosing the metric as $h_{\a\b} = \eta_{\a\b}= (-, +)$, 
we find the canonical momenta and the Hamiltonian as
\begin{mathletters}
\label{can:all}
\beq
P^i &=& \p_\tau X^i - B^i{}_j \p_\s X^j, \quad 
P^a = \p_\tau X^a \label{can:a}\\
H &=& \half \left(P^i + B^i{}_j \p_\s X^j \right)^2 + \half (P^a)^2
+ \half \p_\s X^{\m} \p_\s X_\m \label{can:b}
\eeq
\end{mathletters}
where $a=p+1, \dots, 9$.
The boundary conditions to be imposed are as follows
\beq
\p_\s X^i- B^i{}_j \p_\t X^j = 0, \quad X^a = x^a
\eeq
for $\s=0, \pi$. In terms of the canonical momenta the 
first boundary condition is written by
\beq 
B^i{}_j P^j -M^i{}_j \p_\s X^j=0 \label{bc}
\eeq
where $M^i{}_j = \eta^i{}_j - B^i{}_k B^k{}_j$.
Since the boundary conditions are nontrivial for $X^i$, we will
be concerned only with $X^i$ hereafter. 

We may incorporate the boundary condition Eq.(\ref{bc}) into the canonical
quantization, treating it as a second class constraint. Before going into 
the canonical quantization of the open string in the $D$-brane background,
it may be useful to recall the canonical quantization of the free open
string. The open string is often described as a closed string with an
orbifold condition
\beq \label{orbi}
X^i (\s) = X^i (-\s), \quad P^i(\s) = P^i(-\s).
\eeq
Let us recast this procedure into the canonical quantization.
The Hamiltonian for the free open string is given as
\beq
H = \half \int \frac{d\s}{2\pi} \left[(P^i)^2+(\p_\s X^i)^2\right] 
= \half \sum_{n} \eta_{ij} \left(P^i_n P^j_{-n} + 
n^2 X^i_n X^j_{-n}\right) \nn
\eeq
where $X^i = \sum_n X^i_n e^{in\s}$, $P^j = \sum_n P^j_n e^{-in\s}$.
(It is assumed that appropriate real conditions are imposed on
$X^i_n$ and $P^i_n$.)
The boundary conditions to be imposed on the two ends of open string are
as follows, $\p_\s X^i (0) = \p_\s X^i (\pi) = 0$.
In terms of normal modes these boundary conditions are rewritten as
\beq
\Phi^i_1 = \sum_n n X^i_n=0, \quad
{\bar\Phi}^i_1 = \sum_n n(-1)^n X^i_n =0. 
\eeq
Viewing the boundary conditions as primary constraints, we find that
they generate the secondary constraints
\beq
\Psi^i_1 = \{H, \Phi^i_1 \}_{PB} = \sum_n n P^i_n, \quad
{\bar\Psi}^i_1 &=& \{H, {\bar\Phi}^i_1 \}_{PB} = 
\sum_n n (-1)^n P^i_n. 
\eeq
Here the fundamental Poisson brackets are given by
\beq
\{X^i_n, P^j_m \} = \eta^{ij}\delta_{nm},\quad
\{X^i_n, X^j_m \} = 0, \quad
\{P^i_n, P^j_m \} = 0.
\eeq
The consistency requires that $\Psi^i_1=0$ and ${\bar\Psi}^i_1=0$.
Again in order to impose these secondary constraints consistently 
we should introduce the following constraints
\begin{mathletters}
\label{bc2:all}
\beq
\{H, \Psi^i_1 \}_{PB} &=& -\sum_n n^3 X^i_n=0, \label{bc2:a}\\
\{H, {\bar\Psi}^i_1 \}_{PB} &=& -\sum_n n^3 (-1)^n X^i_n=0. \label{bc2:b}
\eeq
\end{mathletters}

By repetition we get a complete set of constraints
\begin{mathletters}
\label{bc3:all}
\beq 
\Phi^i_m &=& \sum_n n^{2m-1} X^i_n=0, \quad
\Psi^i_m = \sum_n n^{2m-1} P^i_n = 0 \label{bc3:a}\\
{\bar\Phi}^i_m &=& \sum_n n^{2m-1} (-1)^n X^i_n=0, \quad
{\bar\Psi}^i_m = \sum_n n^{2m-1} (-1)^n P^i_n=0 \label{bc3:b}
\eeq
\end{mathletters}
where $m=1, 2, \dots $. 
Since they are of second class, one needs to construct the Dirac bracket
to incorporate them into the canonical quantization. However,
each constraint involves all different modes, the Dirac bracket is 
expected to be complicated. As for this point the 
following simple observation turns out to be very useful.
The set of the constraints, Eq.(\ref{bc3:a}) implies
\beq \label{cons1}
-i\sum_{m=1} \Phi^i_m \frac{(i\s)^{2m-1}}{(2m-1)!} &=&
-i \sum_n \sum_{m=1} X^i_n \frac{(in\s)^{2m-1}}{(2m-1)!} \\
&=& \sum_n X^i_n \sin n\s \nn\\
&=& 0. \nn
\eeq
It follows that 
\beq
\chi^i_n = X^i_n - X^i_{-n} = 0, \quad n=1, 2, \dots, \label{cons2}
\eeq
from Eq.({\ref{cons1}) and
\beq
\int^{2\pi}_{0} \frac{d\s}{\pi} \sin{n\s} \sin{m\s} =
\delta(n-m) - \delta(n+m).
\eeq
We also see that if Eq.(\ref{cons2}) is imposed, the constraint equation,
Eq.(\ref{bc3:a}) $\{\Phi^i_n=0$, $n=1, 2, \dots\}$, holds. 
Thus, two sets of constraints are
equivalent to each other. 
If we apply this procedure to the constraints, Eq.(\ref{bc3:b}) 
$\{{\bar \Phi}^i_m = 0, m = 1, 2, \dots\}$, we get the same result 
as Eq.(\ref{cons2}). 
They do not introduce additional constraints, thus they are redundant.
The same procedure yields that the set of
constraints $\{\Psi^i_n=0$, $n=1, 2, \dots \}$, 
is equivalent to the following set of constraints,
\beq
\varphi^i_n = P^i_n - P^i_{-n} = 0, \quad n=1, 2, \dots \label{cons3}
\eeq
and that the constraints Eq.(\ref{bc3:b}) are redundant.
At a glance we find that these constraints Eqs.(\ref{cons2}, \ref{cons3}) 
are nothing but the orbifold condition Eq.({\ref{orbi}), which introduced
to describe the open string in terms of the closed string. 

Now it becomes an easy task to construct the Dirac brackets.
We evaluate the commutators between the constraints
\beq
\{\chi^i_n, \chi^j_m \}_{PB} = 0, \quad
\{\chi^i_n, \varphi^j_m \}_{PB} = 2\eta^{ij} \delta_{nm}, \quad
\{\varphi^i_n, \varphi^j_m \}_{PB} = 0.
\eeq
Introducing
\beq
C = 2 \left( \begin{array}{cc}
        0 & \eta^{ij} \\
        -\eta^{ij} & 0 \end{array} \right) \otimes I
\eeq
where $I$ is identity matrix, $I_{nm} = \delta_{nm}$,
we construct the Dirac bracket as
\beq
\{A,B\} = \{A,B\}_{PB}- \{A,\phi_M\}_{PB} 
(C^{-1})^{MN} \{\phi_N,B\}_{PB} \label{db1}
\eeq
where $\phi_M =\{ \chi^i_n, \varphi^j_m \}$.
The fundamental Dirac brackets are obtained as
\beq
\{x^i, p^j\}_{PB} = \eta^{ij}, \quad
\{X^i_n, P^j_m\}_{DB} = \half \eta^{ij}
\left(\delta(n-m)+\delta(n+m)\right)
\eeq
where $n$ and $m$ are non-zero integers. All other fundamental
Dirac brackets are vanishing.

\section{Canonical Quantization and $D$-Brane Background}

Now let us return to the canonical quantization of the open string
in the $D$-brane background with a $B$-field. As in the case of the
free open string, the boundary condition may be treated as a primary
constraints. In the presence of $B$-field, we may expand the
canonical string variables in terms of normal modes as
\beq
X^i(\s)= a^i(\s +c)+ \sum_n X^i_n e^{in\s}, \quad 
P^i(\s) = \sum_n P^i_n e^{-in\s} \label{norm}
\eeq
where $c$ is a constant, which will be fixed later.
In the limit of strong $B$-field, the dynamical degrees of freedom
of the open string are mostly encoded by $a^i$ \cite{biga}.
In terms of the normal modes the Hamiltonian and the boundary
conditions Eqs.(\ref{can:all}) read as
\begin{mathletters}
\label{ham:all}
\beq
H &=& \half (p^i + B^i{}_j a^j)^2 + \half (a^i)^2 \nn\\
& & + \sum_{n=1} \left[ \left(P^i_{-n}+ in B^i{}_j X^j_n\right)
\left(P_{in} -in B_{ik} X^k_{-n}\right)+ 
        n^2 X^i_n X_{i -n}\right] \label{ham:a}\\
\Phi^i_0 &=& B^i{}_j \sum_n P^j_n - M^i{}_j (a^j+ i
\sum n X^j_n) = 0, \label{ham:b}\\
{\bar \Phi}^i_0 &=& B^i{}_j \sum_n P^j_n (-1)^n - M^i{}_j
(a^j+ i \sum n X^j_n (-1)^n) = 0. \label{ham:c}
\eeq
\end{mathletters}
Here choose $c=-\pi/2$.
Note that the boundary conditions relate string coordinate variables
$\{X^i_n\}$ to the momentum variables, $\{P^j_n\}$.

Evaluating the commutator between the Hamiltonian and the primary
constraints, 
\beq
\{H, \Phi^i_0 \}_{PB} &=& -\sum_n in P^i_n, 
\eeq
we find the secondary constraints, which are conjugate to the primary
constraints $\Phi^i_0$ 
\beq
\Psi^i_0 = \sum_n n P^i_n = 0.
\eeq
The Dirac procedure requires further that the commutators between
the secondary constraints and the Hamiltonian are vanishing
\beq
\{ H, \Psi^i_0 \}_{PB} =
 -i \sum_n n^2 \left(B^i{}_j P^j_n - in M^i{}_j X^j_n\right)
\eeq
This procedure will be continued until it does not generate additional new
constraints.
By repetition we get a complete set of constraints
\begin{mathletters}
\label{comp:all}
\beq
\Phi^i_m &=& \sum_n n^{2m} 
        (B^i{}_j P^j_n - in M^i{}_j X^j_n) = 0, \label{comp:a}\\
\Psi^i_m &=& \sum_n n^{2m+1} P^i_n=0, \label{comp:b}
\eeq
\end{mathletters}
where $m=0, 1, 2, \dots $. These constraints are of second class.
We also get a set of constraints, which generated by ${\bar \Phi}^i_0$
and its commutator with the Hamiltonian. But as in the case of the
free open string, they are redundant.

Since each constraint involves all different normal modes,
it is desirable to disentangle them to construct the Dirac bracket.
As we observed before, the set of constraints $\{\Psi^i_m=0, m=0, 1, 2,
\dots\}$ is equivalent to 
\beq
\{ \varphi^i_m = P^i_m - P^i_{-m} = 0, 
\quad m = 1, 2, \dots\}. \label{psi}
\eeq
By the similar procedure we disentangle the set of constraints
$\{\Phi^i_m = 0, m=0, 1, \dots\}$.
We are lead to
\beq
\sum_{m=0} \Phi^i_m \frac{(i\s)^{2m}}{(2m)!} &=&
\sum_n (B^i{}_j P^j_n - in M^i{}_j X^j_n) \left(
\sum_{m=0} \frac{(in\s)^{2m}}{(2m)!}\right) -M^i{}_j a^j \nn\\
&=& \sum_n (B^i{}_j P^j_n - in M^i{}_j X^j_n) \cos n\s 
-M^i{}_j a^j \\
&=& 0 \nn
\eeq
It follows from  
\beq
\int^{2\pi}_0  \frac{d\s}{\pi} \cos n\s \cos m\s
= \delta(n-m)+ \delta(n+m), \nn
\eeq
that this set of constraints is
equivalent to the following constraints
\begin{mathletters}
\label{eq:all}
\beq
\chi^i_0 &=& B^i{}_j p^j -M^i{}_j a^j, \label{eq:a}\\
\chi^i_n &=& B^i{}_j(P^j_n + P^j_{-n}) -iM^i{}_j n 
(X^j_n - X^j_{-n}) = 0, \label{eq:b}
\eeq
\end{mathletters}
where $p^i = P^i_0$, and $n = 1, 2, \dots$. 
The first constraint determines $a^i$,
$a^i = (M^{-1} B)^i{}_j p^j$.
Assuming that this solution is used explicitly, we will remove the
constraint $\chi^i_0 = 0$ hereafter. Taking this into account we 
write
\beq
X^i(\s) = (M^{-1}B)^i{}_j p^j (\s+c) +\sum_n X^i_n e^{in\s}.
\eeq
We note that $(M^{-1}B)$ is antisymmetric.

Evaluating the commutators between the constraints, we have
\beq
\{\chi^i_n, \chi^j_m \}_{PB} = 0, \quad
\{\chi^i_n, \varphi^j_m \}_{PB} = -2in M^{ij} \delta_{nm}, \quad
\{\varphi^i_n, \varphi^j_m \}_{PB} = 0. 
\eeq
With this commutator relations we construct
\beq
C = -2i \left( \begin{array}{cc}
        0 & M^{ij} \\
        -M^{ij} & 0 \end{array} \right) \otimes N
\eeq
where $N$ is a diagonal matrix, $(N)_{nm} = n\delta_{nm}$.
The Dirac bracket is defined as Eq.(\ref{db1}) with  
$\{\phi_M\} = \{\chi^1_m, \varphi^i_m\}$ and $C$, which are given by
Eq.(\ref{eq:b}) and Eq.(\ref{psi}) respectively.
The fundamental Dirac brackets are then found be to
\beq
\{x^i, p^j \}_{DB} &=& \eta^{ij}, \quad
\{X^i_n, X^j_m \}_{DB} = \frac{i}{n}(M^{-1}B)^{ij}
\delta(n-m), \label{fund}\\
\{X^i_n, P^j_m  \}_{DB} &=& \half \eta^{ij}\left(
\delta(n-m) +\delta(n+m)\right), \nn
\eeq
where $n$ and $m$ are non-zero integers. Other fundamental brackets
are vanishing.
It is noted that the commutator, $\{X^i_n, X^j_m\}_{DB}$ is modified
due to the background $B$-field, which results in noncommutative
geometry. 

The noncommutativity becomes manifest as we evaluate
the commutator
\beq
\{ X^i(\s), X^j(\s^\prime) \}_{DB} =
- (M^{-1}B)^i{}_j \left[ (\s+\s^\prime+2c)+ \sum_{n\not=0}\frac{1}{n}
\sin n(\s+\s^\prime)\right].
\eeq
Again the consistent choice for $c$ is $c= -\pi/2$. 
The reason will be clear shortly. Making use of
\beq
\sum_{n\not=0}\frac{1}{n} \sin n\theta =
\left\{ \begin{array}{r@{\quad:\quad}l}
\pi -\theta & 0<\theta<2\pi \\
0 & \theta=0, 2\pi \end{array} \right.
\eeq
we have
\beq
\{ X^i(\s), X^j(\s^\prime) \}_{DB} =
\left\{\begin{array}{r@{\quad:\quad}l}
(M^{-1}B)^i{}_j \pi & \s=\s^\prime=0 \\
-(M^{-1}B)^i{}_j \pi & \s=\s^\prime=\pi \\
0 & {\rm otherwise} . \end{array} \right. 
\eeq
Similarly, we find
\beq
\{X^i(\s), P^j(\s^\prime)\}_{DB} &=&
\eta^{ij} \left(1+ \sum_{n\not=0} \cos n\s \cos n\s^\prime \right)\nn \\
\{P^i(\s), P^j(\s^\prime)\}_{DB} &=& 0. \nn
\eeq
These commutator relations agree with those obtained in the work 
of Chu and Ho \cite{ho1,ho2}.

\section{Noncommutative Geometry}
%\section{Solving Constraints}

It is desirable to solve the constraints explicitly if possible. The
constraints are solved explicitly, there is no need for the 
Dirac brackets. Defining 
\begin{mathletters}
\label{sol:all}
\beq
Y^i_n &=& \frac{1}{\sqrt{2}}(X^i_n+X^i_{-n}), \quad
K^i_n = \frac{1}{\sqrt{2}}(P^i_n+P^i_{-n}), \label{sol:a}\\
{\bar Y}^i_n &=& \frac{1}{\sqrt{2}}(X^i_n-X^i_{-n}), \quad
{\bar K}^i_{n} = \frac{1}{\sqrt{2}}(P^i_n-P^i_{-n}), \label{sol:b}
\eeq
\end{mathletters}
where $n = 1, 2, \dots$, we find that the only nontrivial commutation
relations are
\beq
\{Y^i_n, {\bar Y}^j_m \}_{DB} = \frac{1}{n} (M^{-1}B)^{ij} \delta_{nm}, \quad
\{Y^i_n, K^i_m \}_{DB} = \eta^{ij} \delta_{nm}, 
\eeq
and all other commutators are vanishing.
The constraints Eq.(\ref{psi}) and Eq.(\ref{eq:b}) are 
read as
\beq
{\bar Y}^i_n = \frac{1}{in} (M^{-1}B)^i{}_j K^j_n, \quad
{\bar K}^i_n = 0.
\eeq
Using these constraints, we can get rid of 
${\bar Y}^i_n$, and ${\bar K}^i_n$ in
favor of $Y^i_n$ and $K^i_n$, which satisfy the usual commutation
relations.
Accordingly, the Hamiltonian can be written in terms of $Y^i_n$ and 
$K^i_n$ as
\beq
H = \half p^i (M^{-1})_{ij} p^j + \half
\sum_{n=1} \left[ K^i_n (M^{-1})_{ij} K^j_n + 
n^2 Y^i_n (M)_{ij} Y^j_n \right]. \label{hamil}
\eeq
This is precisely the Hamiltonian for a free open string in the
space-time background, of which metric is given by $M_{ij}$.
Thus, the Hamiltonian can be written in terms of the usual commutative
algebra. The noncommutativity arises when we identify the space-time 
coordinates of open strings as
\beq
X^i(\s) =x^i+(M^{-1}B)^i{}_j p^j \left(\s - \frac{\pi}{2}\right)+
\sqrt{2} \sum_{n=1} \left(Y^i_n \cos n\s + \frac{1}{n}
(M^{-1}B)^i{}_j K^j_n \sin n\s \right). \label{coord}
\eeq
The obtained representation for the Hamiltonian, Eq.(\ref{hamil}) and 
the string coordinate variables, Eq.(\ref{coord}) reveals the
nature of the noncommutativity in string theory. The open string
prefers $(Y^i_n, K^i_n)$ as canonical variables while the closed 
string prefers $(X^i_n, P^i_n)$. Thus, in the presence of the 
$D$-brane with a $B$-field, the interaction
between the open and closed strings are expected highly nontrivial.
As we see, the noncommutativity is important not only in the 
zero mode sector but also in all other nonzero mode sectors.
This representation would be useful when we discuss various stringy
noncommutative effects.

In order to compare our results with those of Seiberg and Witten
\cite{seib}, we restore $2\pi\alpha'$, and take $G_{ij} = g_{ij}$.
We also change the signature of the metric on the string worldsheet,
which takes that $h_{\a\b} = (+, +)$ and $B_{ij}$ is replaced with
$iB_{ij}$. Following this prescription, we get the Hamiltonian 
and the boundary conditions as
\begin{mathletters}
\label{new:all}
\beq
H &=& \int \frac{d\s}{2\pi} \left[
\pi\a^\prime g^{ij}(P_i+ i B_{ik} \p_\s X^k)(P_j+iB_{jl}\p_\s X^l) 
-\frac{1}{4\pi\a^\prime} g_{ij} \p_\s X^i \p_\s X^j \right], \label{new:a}\\
0 &=& (Bg^{-1})_{ij}P_j - \frac{i}{(2\pi\a^\prime)^2} 
(G_E)_{ij} \p_\s X^j \quad {\rm for} \,\,\, \s = 0,\,\, \pi
\label{new:b}
\eeq
\end{mathletters}
where $G_E$ is the effective metric seen by the open string \cite{seib}
\beq
(G_E)_{ij} = \left(g - (2\pi\a^\prime)^2 Bg^{-1}B \right)_{ij}. \label{eff}
\eeq
The string coordinate and momentum variables are written by
\beq
X^i(\s) &=& -i (2\pi\a^\prime)^2 \left(G_E^{-1}Bg^{-1}\right)^{ij} p_j
\left(\s - \frac{\pi}{2}\right)
+\sum_n X^i_n e^{in\s}, \\
P^i(\s) &=& \sum_n P^i_n e^{-in\s}. \nn
\eeq
Here we note that
\beq
\left(G_E^{-1}Bg^{-1}\right)^{ij} =
\left(\frac{1}{g+ 2\pi\a^\prime B} B\frac{1}{g- 2\pi\a^\prime B}
\right)^{ij} = -\frac{1}{(2\pi\a^\prime)^2} \theta^{ij}.
\eeq

The fundamental Dirac brackets are given as
\beq
\{X^i_n, X^j_m\}_{DB} &=& -\frac{1}{n} \theta^{ij}\delta(n-m), \nn\\
\{X^i_n, P_{jm}\}_{DB} &=& \half \delta^i{}_j \left( \delta(n-m)+\delta(n+m)\right),\\
\{x^i, p_i\}_{DB} &=& \delta^i{}_j.\nn
\eeq
Other fundamental Dirac brackets are vanishing.
As a concomitant result, we have
\beq
\{ X^i(\s), X^j(\s^\prime) \}_{DB} =
\left\{\begin{array}{r@{\quad:\quad}l}
i \pi \theta^{ij} & \s=\s^\prime=0 \\
- i \pi \theta^{ij} & \s=\s^\prime=\pi \\
0 & {\rm otherwise} . \end{array} \right. 
\eeq
The Hamiltonian and the string coordinate variable are
written in the phase space $(Y^i_n, K^i_n)$ by
\begin{mathletters}
\label{str:all}
\beq
H &=& (2\pi\a^\prime) \half p_i(G_E^{-1})^{ij}p_j
+(2\pi\a^\prime) \sum_{n=1}\left\{ \half K_{in} (G^{-1}_E)^{ij}
K_{jn} -\frac{1}{(2\pi\a^\prime)^2} \frac{n^2}{2} Y^i_n (G_E)_{ij} Y^j_n 
\right\}  \label{str:a}\\
X^i(\s) &=& x^i+i \theta^{ij}p_j \left(\s - \frac{\pi}{2}\right)+
\sqrt{2} \sum_{n=1} \left(Y^i_n \cos n\s + \frac{i}{n}
\theta^{ij} K_{jn} \sin n\s \right). \label{str:b}
\eeq
\end{mathletters}
Thus, the obtained Hamiltonian is precisely the Hamiltonian for
a free open string in space-time with the metric given by $(G_E)_{ij}$
as we may expect. Concomitantly the spectrum of the open
string in the $D$-brane background is determined by the effective metric
$(G_E)_{ij}$. As we mentioned before, it is convenient to employ basis
$\{ |Y^i(\s)> \}$ to describe the open string, interacting with
the $D$-brane, while the usual basis $\{ |X^i(\s)> \}$ is more
suitable for the closed string. Note that the eigenstate of $X^i(\s)$, 
$|X^i(\s)>$ can be constructed as a coherent state in $\{ |Y^i(\s)> \}$.
It is quite similar to the lowest Landau level state. More detailed
discussion on this point will be given somewhere else \cite{tj}.
In ref.\cite{seib}, Siberg and Witten discuss the zero slope limit, 
where $ \a^\prime \sim \epsilon^{\half} \rightarrow 0, \,\,
g_{ij} \sim \epsilon \rightarrow 0 $
while keeping $B$, $G_E$ and $\theta$ fixed.
The zero slope limit does not alter the noncommutative structure,
but it makes the potential term dominant in the Hamiltonian
as in the lowest Landau level. As we see even in the zero slope limit
the nonzero mode sectors contribute to the noncommutative 
$D$-brane dynamics as well as
the zero mode sector. The nonzero mode sectors would be important
to understand some stringy effects in the noncommutative geometry.

\section{Concluding Remarks}

We conclude this paper with a few remarks.
We find that the dynamics of $D$-brane with a $B$ field can be
understood in the framework of the canonical quantization.
In the presence of the $B$ field we have a mixed boundary
condition, which generates an infinite number of secondary
second class constraints.
The set of the second class constraints is shown to be equivalent to an
orbifold condition, which is a generalization of the simple
one introduced when the free open string is described in
terms of the closed string.
The best way to deal with the second class constraints is to solve
them explicitly. Indeed we can solve the constraints explicitly
without difficulty and get a simple Hamiltonian for the
open string in the $D$-brane background. The Hamiltonian is found
to be a free Hamiltonian for an open string in space-time with
the metric $G_E$ Eq.(\ref{eff}). Noncommutativity arises as the
orbifold condition effectively reduces the phase space for the
string by half. Eq.(\ref{str:all}) reveals the nature of 
noncommutativity in string theory transparently. The present work
may serve as stepping stone leading us to various directions.
The canonical analysis carried out in the present paper may enable
us to construct the second quantized theory for the open string in the
$D$-brane background, which would be an appropriate generalization
of the earlier work on the open string by Witten \cite{witten}.
We may also apply the same canonical quantization procedure
to the open string attached to the multi-$D$-branes or to two
different types of $D$-branes. Work along this direction
may improve our understanding of the $AdS/CFT$ correspondence \cite{mal}
and the black hole physics in string theory. In the due course
one may attempt to derive the (non-Abelian) noncommutative 
Dirac-Born-Infeld effective action \cite{seib,tsey} for the $D$-brane,
which remains to be an outstanding open problem.

\section*{Acknowledgement}
This work was supported in part by the Basic Science Research 
Institute Program, Ministry of Education of Korea (BSRI-98-2401). 
Part of the work was done during his visit to PIMS (Canada) and 
KIAS (Korea).

\end{document}